\documentclass{ws-p8-50x6-00}

\newcommand{\etal}{{\em et al.}}

\begin{document}

\title{Probing Halo and Molecular States in Light, Neutron-Rich Nuclei}

\author{N.A.~Orr}

\address{LPC-ISMRa, Bd Mar\'echal Juin, 14050 Caen Cedex, France\\ 
e-mail: orr@caelav.in2p3.fr}

\maketitle

\abstracts{Selected topics on halo and molecular states in light, 
neutron-rich nuclei are discussed.  In particular, work 
on $x\alpha$:X$n$ structures is briefly reviewed.  The use of proton radiative
capture as a probe of clustering is also presented through the
example provided by the $^6$He(p,$\gamma$) reaction.  
}

\section{Introduction}

Until relatively recently cluster studies
have been confined to the region encompassing the line of beta 
stability.  
As clustering is expected to manifest itself most
strongly near thresholds\cite{Ikeda}, 
exotic structures might
be expected to form in the more weakly bound systems found in the vicinity of 
the driplines.  Perhaps the most striking example of this are the halo nuclei.

In the present paper two topics are addressed.  Firstly a
brief review is given of the status of work on $x\alpha$:Xn 
molecular states.  Secondly radiative capture of protons is investigated as 
a probe of clustering through the example provided by the 
reaction $^6$He(p,$\gamma$) at 40~MeV.

\section{Molecular States}

Owing to the strongly bound character of the $^{4}$He nucleus
and the weakness of the $\alpha$-$\alpha$ interaction 
the $\alpha$-particle plays an important
r\^ole in the structure of light $\alpha$-conjugate nuclei. 
Whilst an excess of
neutrons (or protons) might na\"ively be expected to dilute any underlying $\alpha$-cluster
structures, 
theoretical\cite{Sey81,AMD} and recent experimental work\cite{vonO,Fre99}
indicate that molecular-type structures such as $\alpha$-chains ``bound'' by valence
nucleons also occur.
The appearance of such cluster structures is well illustrated, as discussed here, 
by the beryllium isotopes,
for which the $\alpha$-$\alpha$ system may be regarded as the basis.  

As described at this symposium a variety of theoretical models
exist.  For example, the Molecular-Orbital Model (MO)\cite{Sey81}, in which valence nucleons
are added to the single-particle orbits arising from the two-centre potential, provides a
conceptually appealing framework within which to describe the properties 
of these nuclei. Moreover these orbits may be
interpreted as the analogues of the $\sigma$ and $\pi$-orbitals associated with the covalent
binding of atomic molecules. The development of 
fully fledged 
Antisymmeterised Molecular Dynamics calculations (AMD), as discussed by 
Kanada En'yo in 
her contribution to YKIS01, is of particular 
interest as the
nucleus is modelled without any a priori imposition of an underlying
cluster structure.  

From an experimental perspective systematic
evidence for the existence of dimers in $^{9-11}$Be and $^{9-11}$B has been 
compiled\cite{vonO}.
In the case of $^{9}$Be, for example, 
many facets of the level structure may be understood
in terms of a three-body $\alpha$:n:$\alpha$ molecular structure.  
In particular,
the rotational bands based on the ground and low-lying states exhibit large
deformations consistent with the associated molecular configurations.

\begin{figure}
\centerline{\epsfig{file=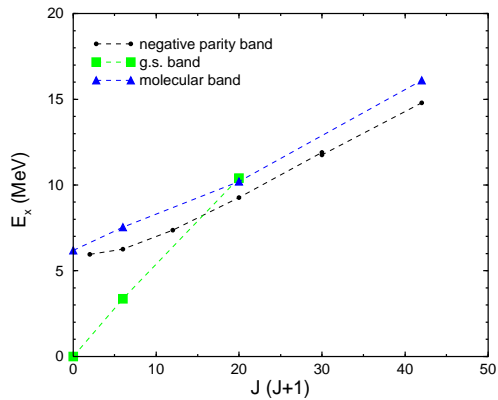,width=6.5cm}}
\caption{Spin-energy systematics for states observed in $^{10}$Be 
(from ref.\protect\cite{Fre00}).
The trajectories of the postulated positive and negative parity molecular bands are 
indicated.}
\vspace*{-5mm}
\end{figure}

In the case of $^{10}$Be, the experimental evidence for molecular configurations
is rather less well documented.  Beyond the 
established 0$^+_2$, 2$^+_2$ and 1$^-_1$ -- 4$^-_1$ states,
the locations of the J=5\cite{vonO} and 6 members of the of the negative 
parity band, as well as the 
J=4 and 6 members of the positive parity band have been postulated following 
studies of the $\alpha$-$^{6}$He breakup of $^{10}$Be$^*$\cite{Soic,Fre00}.
Recently, Curtis {\em et al.} have succeeded in determining the
spins of the levels at 9.56 (J=2) and 10.15~MeV (J=3)\cite{Cur01}
As displayed in Fig. 1, the spin-energy trajectories for the bands based 
on the
0$^+_2$ and 1$^-_1$ states at $\sim$6~MeV are consistent with 
large deformations ($\hbar^2/2\Im$$\simeq$0.23~MeV) 
as expected for molecular-like $\alpha$:2n:$\alpha$
structures.  
Theoretical support may be found in AMD
calculations, whereby 
well developed $\alpha$:2n:$\alpha$ configurations are predicted for the 
0$^+_2$ and 1$^-_1$ bands\cite{AMD}.  
Furthermore, in extended MO model calculations the 0$^+_2$ state appears to 
be well characterised by valence neutrons
occupying the $\sigma$-orbital\cite{Ita00}.

Given the existence of such molecular-type structures in $^{10}$Be, the question naturally
arises as to the existence of similar structures in more neutron-rich systems.
In this context we investigated the
dripline nucleus $^{12}$Be via
inelastic scattering at 35~MeV/nucleon.  
Evidence was found in these measurements for the breakup into 
$^{6}$He+$^{6}$He and $\alpha$+$^{8}$He 
of states
(J=4, 6, 8)
in the range E$_x$=10-20~MeV which exhibit spin-energy systematics
characteristic of a rotational band\cite{Fre99,Fre00}.
Moreover the inferred momenta of inertia -- $\hbar^2/2\Im$=0.15$\pm$0.04~MeV -- 
and bandhead 
energy (10.8$\pm$1.8~MeV)
are consistent with the cluster decay
of a molecular structure which may be associated with $\alpha$:4n:$\alpha$
configurations.
As reported at this symposium by Saito, an experiment using two neutron 
removal from an energetic $^{14}$Be beam has 
recently uncovered a probable 0$^+$ state some 1.7~MeV (E$_x$=11.8~MeV)
above the $^6$He+$^6$He breakup threshold, as well as weaker evidence
for the levels observed in our work.  
The location of this new level conforms reasonably well to the 
spin-energy systematics established in our original study and suggests, interestingly, 
that the
bandhead lies above the $^6$He+$^6$He threshold.

The theoretical investigation of molecular configurations in $^{12}$Be represents a
more challenging venture than the lighter mass Be isotopes.  Nevertheless,
efforts are underway, as evidenced by the contribution of Ito to this symposium
and recent papers by Itagaki {\em et al.}\cite{Ita00} and 
Descouvemont and Baye\cite{Des01}.
 
As suggested by von Oertzen\cite{vonO} and more recently by Itagaki {\em et al.}\cite{Ita01}, 
the neutron-rich C isotopes may be expected to exhibit 3$\alpha$:Xn cluster structures.  
In this context, we have attempted to observe such states in the  
$^{12}$C($^{16}$C,$^{16}$C$^*$$\rightarrow$$^{10,12}$Be$^*$+$^{6,4}$He) reaction 
at 35~MeV/nucleon\cite{Lea01}.  Careful analysis of the associated $^{6}$He+$^{6}$He+$\alpha$
and $^{8}$He+2$\alpha$ fragment coincidences could, 
however, only put an
upper limit of some 2$\mu$b on the yield to states in these decay 
channels\footnote{An upper limit of 30$\mu$b could be put on the 
$^{10,12}$Be$_{gs}$+$^{6,4}$He
decay channels.}.  The
inability to access such states by inelastic scattering may arise from a much
smaller overlap between the $^{16}$C ground state and the cluster states than in the
case of $^{12}$Be.

\section{Radiative proton capture on $^6$He}

A recent 
investigation of coherent bremsstrahlung production in the
reaction $\alpha$(p,$\gamma$) at 50~MeV has demonstrated, as described in 
the accompanying contribution by Herbert Loehner, that the high-energy
photon spectrum is dominated by capture to form $^5$Li\cite{Hoe00}. This
result motivated us to 
extend the technique to probe clustering in more exotic systems\cite{Sau00}.  As 
a first test $^6$He was chosen owing to the relatively high beam intensities 
available
and the fact that structurally it is the most well established two-neutron halo
nucleus. Given a
proton wavelength of 0.7~fm at 40~MeV, 
direct capture might be observed, as a quasi-free process, on the constituents 
($\alpha$+n+n) of $^6$He
in addition to capture into $^7$Li. Moreover, the different quasi-free capture
(QFC) processes would lead to different $E_\gamma$ in the range 20--40~MeV. 


Experimentally, a 40~MeV/nucleon $^6$He beam (5$\times$10$^5$~pps) was employed
to bombard a solid hydrogen target (95~mg/cm$^2$). The 
different charged reaction products  
were identified and momentum analysed using the SPEG
spectrometer. The photons were detected using 74 elements of the 
``Ch\^ateau de Cristal'' BaF$_2$ array,
with a  
total efficiency of some 70\%.  Further details including the 
analysis techniques may be found in ref.\cite{Sau00}. 
 
\begin{figure}[tb]
 \begin{center}
  \mbox{\psfig{file=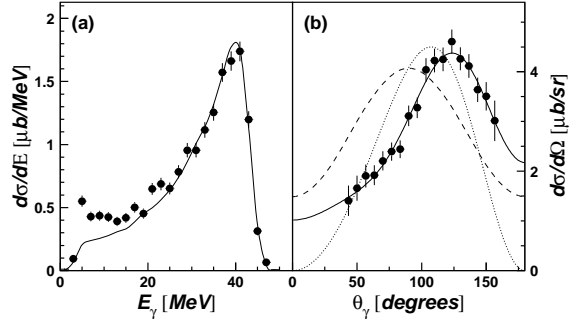,width=7.5cm}}
 \end{center}
 \caption{Energy (a) and angular distributions (b) in the $^6$He+p c.m.\ for
photons in coincidence with $^7$Li. The solid line in (a) is the response of
the Ch\^ateau to $E_\gamma=42$~MeV. The lines in (b) are a classical
electrodynamics calculation\protect\cite{Hoe99} (dotted), a cluster 
model\protect\cite{Sau00,Des95} (dashed),
both normalized to the data, and a Legendre polynomial fit\protect\cite{Wel82} (solid).}
\vspace*{-6mm}
\end{figure}
 

Turning to the experimental observations, the reaction
$^6$He(p,$\gamma$)$^7$Li is unambiguously identified by the $\gamma$-rays in
coincidence with $^7$Li (Fig.~2). In particular, the photon energy
spectrum, as well as the $^7$Li momentum\cite{Sau00}, is well described
assuming a $\gamma$-ray line at 42~MeV. 
The energy difference between the two particle-stable states of $^7$Li -- the 
g.s. and the first
excited state at 0.48~MeV) -- is too small for them to be
distinguished in this experiment. 
A total cross section of $\sigma=35\pm2~\mu$b was deduced. 
 
The $^6$He(p,$\gamma$)$^7$Li cross section has been calculated using a
microscopic cluster model\cite{Des95}. 
At 40~MeV, a cross section
of $\sigma=59~\mu$b was found, with 15$\mu$b going to the g.s.\ and
44$\mu$b to the first excited state\cite{Sau00}. 
The calculation was restricted to the dominant E1 multipolarity, thus leading to an
angular distribution symmetric about $90^\circ$ (Fig.~2b). The cross
section to the g.s. can also be estimated from photodisintegration measurements\cite{Sen85} 
via
detailed balance considerations and is $9.6\pm0.4~\mu$b. Given the predicted
relative populations of the ground and first excited state, a total capture
cross section of $\sigma\sim38~\mu$b is obtained, in agreement with the value
measured here.
  

QFC was investigated by searching for $\gamma$-rays in coincidence with
fragments lighter than $^7$Li. The corresponding energy spectra
(Fig.~3a,c,e) do indeed exhibit peaks below 42~MeV. In order to
establish the origin of these fragment-$\gamma$ coincidences, QFC processes on
the subsystems of $^6$He have been modelled as follows. The $^6$He projectile
was considered as a cluster ($A$) plus spectator ($a$) system in which each
component has an intrinsic momentum distribution, the corresponding energy
$E_A+E_a-m_{^6\rm{He}}$ being taken into account in the total available energy.
The reaction may be denoted as $a$+$A$(p,$\gamma$)$B$+$a$, and the $\gamma$-ray
angular distribution is assumed to be that given by the charge asymmetry of the
entrance channel\cite{Hoe99}. The intrinsic momentum distribution of
all the clusters was taken to be Gaussian in form (${\rm{FWHM}}=80$~MeV/$c$).
 In order to explore the possibility that FSI may occur in the exit channel
between the spectator, $a$, and the capture fragment, $B$, an extended version
of the QFC calculation was developed\cite{Sau00}. Here the energy in the system
$B$+$a$ is treated as an excitation in the continuum of $^7$Li, which decays in flight.
 
In the case of $^6$Li-$\gamma$ coincidences, two lines were observed
(Fig.~3a) at 30 and 3.56~MeV corresponding to the
formation of $^6$Li and the decay of the second excited state.
It was estimated that $^6$Li is
formed almost exclusively ($96^{+4}_{-24}\%$) in the 3.56~MeV excited state.
The deduced cross section was $\sigma=3.5\pm1.3~\mu$b. The lines in
Fig.~3a,b corresponds to QFC on $^5$He into
$^6$Li$^*$(3.56~MeV). The $\gamma$-ray energy spectrum is well described, whilst
the $^6$Li momentum distribution requires inclusion of $^6$Li-n
FSI. 

\begin{figure}[tb]
 \begin{center}
  \mbox{\psfig{file=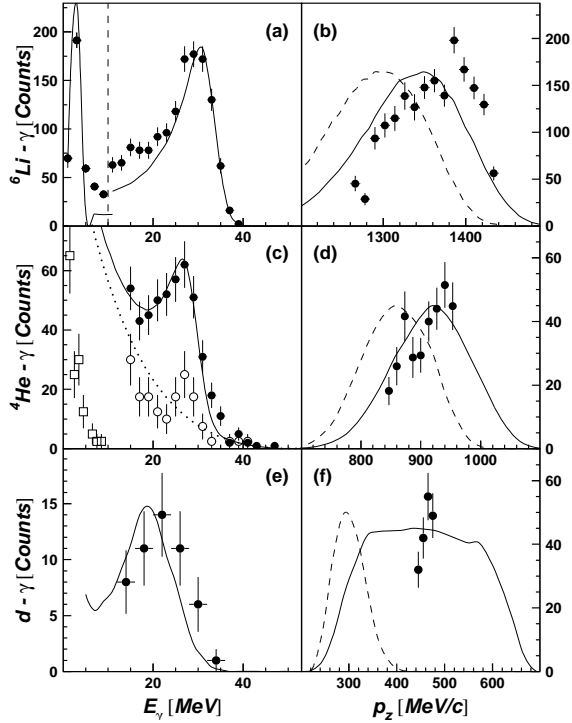,width=7.5cm}}
 \end{center}
 \caption{Gamma-ray energy spectrum in the $^6$He+p c.m.\ and momentum
distribution of the coincident fragment for $^6$Li (upper), $\alpha$ particles
(middle) and deuterons (lower panel). The lines correspond to calculations of
QFC on the $^5$He cluster, the $\alpha$ core and one halo neutron,
respectively, on the right with/without (solid/dashed) fragment FSI (see text).
The distribution in (a) was divided by 3 below 10~MeV, and the open symbols in
(c) are from an analysis investigating the role of the neutron background (see
ref.\protect\cite{Sau00}).}
\vspace*{-6mm}
\end{figure}
 

Evidence for QFC on the $\alpha$ core, whereby the two
halo neutrons would behave as spectators,  has also been searched for. 
The photon spectrum should resemble
that observed for the $\alpha$+p reaction\cite{Hoe00}. Indeed such a
$\gamma$-ray energy spectrum (Fig.~3c) was observed.
The background, however, arising from
$^6$He breakup, in which the $\alpha$ particle is detected in SPEG and the halo
neutrons interact with the forward-angle detectors of the Ch\^ateau, is
significant. In order to minimise this background, only the backward-angle
detectors ($\theta>110^\circ$) of the Ch\^ateau were used in the analysis.
The $\gamma$-ray spectrum under this condition exhibits two components: a peak
at $E_\gamma=27$ MeV and a $1/E_\gamma$ continuum similar to coherent
$\alpha$+p bremsstrahlung\cite{Hoe00}.
 
Simulations indicate, however, that some back-scattered neutrons remain from
breakup, which would also lead to a continuous component
with a $1/E$ type spectrum in the Ch\^ateau\cite{Sau00}. 
A single background component of this form (dotted line, Fig.~3c) was 
therefore added to the QFC process
$\alpha$(p,$\gamma$)$^5$Li. The photon energy spectrum is thus well described,
as is the momentum distribution of the $\alpha$ particle. The cross section was
estimated to be $\sigma=4\pm1~\mu$b. Additional support for this interpretation 
is found in
$\alpha$-$\gamma$-n coincidences, for which 30 events were 
observed\cite{Sau00} (open symbols, Fig.~3c).


Finally, d-$\gamma$ coincidences presenting a peak in the $\gamma$-ray energy
spectrum, at $E_\gamma\simeq$21~MeV, were also observed (Fig.~3e). 
The relatively low statistics arose from
the limited acceptances of the spectrometer for deuterons (Fig.~3f).
The predictions for n(p,$\gamma$)d QFC on a
halo neutron present a peak at 19~MeV (Fig.~3e) -- the small shift may
be attributable to the strong kinematic correlation between the deuteron
momentum and the photon energy, as the detection of only a small fraction of
the deuterons is
predicted\cite{Sau00}. As such no reliable estimate of the
cross section was possible.
 
There are additional QFC channels, 2n(p,$\gamma$)t and t(p,$\gamma$)$\alpha$, 
that could have been observed with finite efficiency but 
were not\cite{Sau00}. Perhaps the most interesting is QFC on the two halo
neutrons. In the case of $^6$He, several theoretical models predict the
coexistence of two configurations in the g.s. wave function: the so-called
``di-neutron'' and ``cigar'' configurations\cite{Zhu93}. Here one might expect
that the different admixtures of these could be probed by the relative strength
of the n,2n(p,$\gamma$)d,t QFC processes, whereby the corresponding free cross
sections at 40~MeV, obtained from detailed balance considerations, are
comparable: 9.6$\mu$b\cite{Ahr74} and 9.8$\mu$b\cite{Fau80}, respectively.
However, events registered in the Ch\^ateau in coincidence with tritons in SPEG
have energies below 10~MeV, whereas the 2n(p,$\gamma$)t reaction should produce
photons with $E_\gamma\approx32$~MeV. 
  
As described above, the QFC with fragment FSI model describes well the observed
monoenergetic $\gamma$-rays, as well as the momentum distribution of
the capture fragment ($B$). The $\gamma$-ray lines are associated with
specific energy distributions for the fragments in the exit channel.
Therefore, such a process will exhibit the same kinematics as capture into
continuum states above the corresponding threshold,
$^6$He(p,$\gamma$)$^7$Li$^*$$\rightarrow$$B$+$a$, provided that the equivalent
region of the continuum is populated\cite{Sau00}. If, however, all the
final states observed here were the result of radiative capture into $^7$Li,
capture via the non-resonant continuum in $^7$Li might well be expected to
occur\cite{Sid86}. This would lead to a continuous component to the
$\gamma$-ray energy spectra. Moreover, events corresponding to
$E_{^7\rm{Li}^*}=0.5$--10~MeV have not been observed in either t-$\gamma$
coincidences or $\alpha$-$\gamma$ coincidences with $E_\gamma=32$--42~MeV, nor
has the decay into $\alpha$+t for $E_{^7\rm{Li}^*}>10$~MeV. Within the picture
of QFC on clusters, this is simply explained by the absence of the
2n(p,$\gamma$)t and t(p,$\gamma$)$\alpha$ QFC processes for the $^4$He-2n
and t-t configurations, respectively, indicating that
$^4$He-n-n is the dominant configuration in $^6$He.  This is in agreement with the
a recent neutron-neutron interferometry measurement
we have performed ref.\cite{FMM00}.
 
\section{Conclusions}

An aper\c{c}u of $x\alpha$:X$n$ clustering has been presented and some examples 
from the neutron-rich Be isotopes discussed.
In the very near future, transfer reactions
using the combination of low-energy radioactive beams such as $^{6,8}$He and targets 
($^{6,7}$Li, $^9$Be, $^{12}$C) presenting $\alpha$-clustering will be investigated 
as a tool to access molecular-type states in neutron-rich
nuclei.  Additionally, it is hoped that partial decay widths may be determined.

Radiative proton capture has been explored as a probe of clustering in the 
ground states of nuclei far from stability through the example of a measurement 
on the halo nucleus $^6$He. In addition to $^6$He(p,$\gamma$)$^7$Li,
evidence for QFC on $^5$He, $\alpha$ and n was found. 
Of particular importance was the observation of events which
correspond to the previously measured $\alpha$(p,$\gamma$) reaction, as well as
the non-observation of capture on a di-neutron. Theoretically, 
models need to be developed to describe capture on the constituent
clusters of exotic nuclei and, for comparison, capture on the projectile into
unbound final states.

\section*{Acknowledgments}

I would like to underline the leading r\^oles played by
Martin Freer (molecular states), Emmanuel Sauvan and 
Miguel Marqu\'es (radiative capture).  It is a
pleasure also to thank the members of the E281 and E302 collaborations for their efforts 
and
the SPEG crew for developing the hydrogen target.

\end{document}